# "GRAVITATIONAL MASS" OF INFORMATION?[1]

L. B. KISH[2],
*Department of Electrical and Computer Engineering, Texas A&M University, College Station, TX 77843-3128, USA*

We hypothesize possible new types of forces that would be the result of new types of interactions, static and a slow transient, between objects with related information contents (pattern). Such mechanism could make material composition dependence claimed by Fishbach, *et al.* in Eotvos type experiments plausible. We carried out experiments by using a high-resolution scale with the following memories: USB-2 flash drives (1-16GB), DVD and CD disks to determine if such an interaction exist/detectable with a scale resolution of 10 microgram with these test objects. We applied zero information, white noise and 1/f noise type data. Due to the non-reproducible changes of static weight of these memories after changing the information in them, we have not been able to clarify the existence of a reproducible static force between the memory and its environment even though the variations of static weight could possibly be the manifestation of the new type of interaction with changing environment. Interaction between two memories containing the same information was not detected at the given weight resolution. The hypothesis of slow transient interaction is more in line with the observations provided the observed effects are not artifacts. Writing or deleting the information in any of these devices causes peculiar negative weight transients, up to milligrams (mass fraction around $10^{-5}$), which is followed by various types of relaxation processes. These relaxations have significantly different dynamics compared to transients observed during cooling after stationary external heating. Interestingly, a USB-1 MP3 player has also developed comparable transient mass loss during playing music, even though its information content was not changed, and its power dissipation and warming-up was much less than that of the USB-2 flash drives. A classical interpretation of the negative weight transients could be absorbed water in hygroscopic components though comparison of relaxation time constants with air humidity data does not support an obvious explanation. Another classical interpretation with certain contribution is the lifting Bernoulli force caused by the circulation due to convection of the warm air. However, in this case all observed time constants with a device should have been the same unless some hidden parameter causes the observed variations. Thus further studies are warranted to clarify if there is indeed a new force, which is showing up as negative mass at weight measurement when high-density structural information is changed or read out (measured). We estimate that, if the information-based interaction is real, the weight of bodies could be reduced and floating engines could be achieved with several orders of magnitude greater data handling capacities than today's values at the same mass.

## 1. Introduction

Measured anomalies in the gravitation constant inspired several theories of new short-range forces; such the fifth force initiative of Fischbach and coworkers, even though the interpretation of experimental data is somewhat controversial [1-8]. In this Current Opinion, we hypothesize possible new types of forces that would be the result of new types of interactions between objects with related information contents (pattern). Such interaction would potentially be able to cause variable force deviations in gravitational measurements depending on the structural information carried by the test object and the environmental patterns in the vicinity. Such mechanism could make material composition dependence claimed by Fischbach, *et al.* [2] in Eotvos type experiments plausible. We carried out experiments by using a high-resolution scale with USB-2 flash drives (1-16GB), DVD and CD disks to determine if such an interaction exist/detectable with a scale resolution of 10 microgram with these test objects. We applied zero information, white noise and 1/f noise type data. Due to the non-reproducible changes of static weight of these memories after changing the information in them, we have not been able to clarify the existence of a reproducible static force between the memory and its environment even though the large variations of static weight could possibly be the manifestation of the new type of interaction with changing environment. Interaction between the memories with the same information was not detected at the given weight resolution. The hypothesis of slow transient interaction is more in line with the observations provided the observed effects are not artifacts. Writing or deleting the information in any of these devices caused peculiar negative weight transients, up to milligrams (mass fraction of about $10^{-5}$), which was followed by various types of relaxation processes. These relaxations have significantly different dynamics compared to transients observed during cooling after stationary external heating. Interestingly, a USB-1 MP3 player has also suffered comparable mass loss during playing music, even though its information content was not changed, and its power dissipation and warming-up was much less than that of the USB-2 flash drives. A classical interpretation of the negative weight transients could be absorbed water in hygroscopic components though comparison of relaxation time constants with air humidity data does not support an obvious explanation. Another classical interpretation with certain contribution is the lifting Bernoulli force caused by the circulation due to convection of the warm air. However, in this case all observed time constant should have been the same unless some hidden parameter causes the observed variations. Thus further studies are warranted to clarify if there is a new force, which is showing up as negative mass at weight measurement when high-density structural information is changed or read out (measured). Similar measurements should be performed with other types of scales in a possibly more controlled environment. However we estimate that, if the information-based interaction is real, the weight of bodies could be reduced and floating engines could be achieved with about ten to hundred thousand times greater data handling performance than today's values at the same mass.

[1] http://arxiv.org/abs/0711.1197
[2] until 1999: L.B. Kiss

Note, by no means this paper should be considered as a thorough study with firm conclusions. This is just an exploration of new questions that may lead to a new field of study. The author made an effort to clarify the situations, as much as he could with the available limited resources and time effort, however he still may have been ignorant about known aspects in related fields such as the given environment modifying the convection-circulation in the confined scale space or the behavior of hygroscopic materials; which both may lead to variations in the observed relaxation times if all transient effects have classical thermal origin. However, we believe that the current results justify further studies with significantly greater resources and efforts (such as precision gravitation experiments).

## 2. Hypotheses and considerations

### 2.1. *The hypothesized interaction types*

#### 2.1.1. *The core hypothesis: a static interaction between structures with rich information content*

Our core hypothesis to explain Fischbach's and his coworkers' claims [2] about the anomalies in the Eotvos experiments is a new type of static interaction between two objects with related information contents (pattern) where a quantity $I_{1,2}$ describes the correlation of the information patterns (structural information) carried by the two bodies. Generally, such interaction does not have to be based on energy exchange of the systems, it can be for example interaction of probabilities such in classical teleportation theories [9], however in the present paper we suppose an interaction based on energy exchange.

For example, in digital patterns the *mutual information* of these patterns may be the proper quantity to use here. For discrete variables:

$$I_{1,2} = \sum_{x_i = H,L} \sum_{y_j = H,L} p_{1,2}(x_i, y_j) \log\left[\frac{p_{1,2}(x_i, y_j)}{p_1(x_i) p_2(y_j)}\right] \quad (1)$$

where $p_1(x_i)$ is the probability density function that the i-th element $x_i$ of *pattern-1* belonging to *body-1* has a given value (high or low, *x=H* or *x=L*, respectively); similarly $p_2(y_j)$ is the probability density function that the j-th element $y_j$ of *pattern-2* belonging to *body-2* is a given value (*y=H* or *y=L*, respectively); and $p_{1,2}(x,y)$ is the joint probability function of these situations. The square of $I_{1,2}$ has properties resembling to gravitation because by multiplying the size and mass $M_1$ and $M_2$ of the bodies while repeating the same specific patterns in them will cause $I_{1,2}^2$ to grow proportionally to the product $M_1 M_2$ , that is, $I_{1,2}^2$ scales proportionally with the gravitational force between the two bodies.

Because it is unknown if such interaction exists and if it follows the characteristic of gravitation mentioned above, generally we can hypothesize that the total potential energy $V$ of the interaction between two electrically neutral bodies may be written as

$$V = V_{grav} + V_{inf} = -\gamma \frac{M_2 M_2}{R} + V_{inf}(I_{1,2}, P, R) \quad (2)$$

where $V_{grav}$ is the potential energy of gravitational interaction; $V_{inf}$ is the potential energy due to the interaction of the structural information in the two bodies; $\gamma$ is the gravitational constant; and $R$ is the distance (of point-like bodies). $V_{inf}$ is a presently unknown function of the correlation $I_{1,2}$ of the structural information of the two bodies; the distance $R$ ; and the structural parameters $P$ which may be a list or function of parameters in this equation. In digital patterns, $P$ may characterize the energy (or energy change, or barrier height) relevant to the different bit states; the spacing or spatial distribution between bits; etc.

From Eq. 2, the force between the two interacting bodies is:

$$F = -\frac{dV}{dR} = -\gamma\frac{M_1 M_2}{R^2} - \frac{dV_{inf}(I_{1,2}, P, R)}{dR} \tag{3}$$

As an illustration, we may suppose that, within the range limit of interaction, the interaction potential is simply given as:

$$V_{inf} = \frac{\eta P}{R} I_{1,2}^2 \tag{4}$$

where $\eta$ is the information-interaction constant; $I_{1,2}$ is the mutual information defined by Eq. 1; and $P$ is a constant depending on the physical parameters of the interacting structures. If $\eta$ and $P$ are positive then, according to Eq. 3, the force between the two (point-like) bodies is:

$$F = -\frac{dV}{dR} = -\gamma\frac{M_1 M_2}{R^2} + \frac{\eta P}{R^2} I_{1,2}^2 \quad , \tag{5}$$

which represent a repulsive "anti-gravitation" force scaling with distance in the same way as gravitation. The presence of this kind of force could be tested by measurement of the force between bodies at different information contents of one of the bodies or both of the bodies. Alternatively, the ratio of gravitational mass and inertial mass at different information fillings can also be an indicator.

### 2.1.2. *Additional hypothesis: transient force indicated by experiments (can be artifacts)*

Due to our surprising experimental results (see Sections 3,4), and the present lack of a full interpretation of the data in a classical way; until a more standard/plausible explanation based on convection of warm air, condensed water in hygroscopic parts of the test bodies or other unknown mechanism is confirmed; we can hypothesize also a transient force term which is generated by information change and it rises and decays slowly:

$$F_{trans} = F\left\{\left(\frac{dI_{1,2}}{dt}\right)^n, P, R, t\right\} \tag{6}$$

where $n \ge 2$ is an even number and the details of the slow relaxation function $F$ are unknown except that, due to information change/flow, it produces a repulsive ("anti-gravitation") type force with long relaxation times. Note, such a phenomenon may result in an interaction force even if the structural information were not changed but *measured/processed* by a classical physical device. Then the interaction would occur through the device depending on how large fraction of the information is stored in the device at any instance of time. Alternatively, if the transient interaction were a slow interaction with "time-integrative" fashion, such as relaxation processes in disordered media, the "memory" storing the information would be inherently included in the interaction mechanism.

It is important to note that any of the above mentioned effects would result in peculiar phenomena that are unknown to the present state-of-the-art of physics, such as an excess energy need for a system to create certain information content (possibly with certain timing) that are within the range of significant static interaction. However, due to the weakness of such forces and due to the relatively small size of information we can store today, such an effect may have stayed below the limits of observations at everyday situations.

## **2.2.** *Energy conservation law*

Let us suppose that the energy conservation law holds also for these interaction. Then it is easy to see that the *static* interaction described by Eq. 4 and by the last term of Eq. 5 cannot be measured by our present equipment of 10 microgram weight resolution (force resolution $\approx 10^{-7}\ N$ ) and information storage capacity (less then 16GB) of test devices unless the range dependence ($R$-dependence in Eqs. 4 and 5) of the interaction is something very different.

According to energy conservation law, the invested electrical energy $E_e$ for making the information change in the structure must be greater or equal to the created interaction potential:

$$E_e > V_{inf} = \frac{\eta P}{R} I_{1,2}^2 \qquad (7)$$

$$\frac{E_e R}{I_{1,2}^2} > \eta P \qquad (8)$$

$$\frac{\eta P}{R^2} I_{1,2}^2 > 10^{-7} \qquad (9)$$

$$\frac{E_e R}{I_{1,2}^2} > \eta P > \frac{10^{-7}}{I_{1,2}^2} R^2 \qquad (10)$$

$$E_e > 10^{-7} R \quad \text{or} \quad R < 10^7 E_e \qquad (11)$$

$$E_e \approx 1W * 1000s = 100J \quad \Rightarrow \quad R < 10^9 m \tag{12}$$

$$E_e \approx 2000kT * 10^{11} \approx 10^{-6} J \quad \Rightarrow \quad R < 10m \tag{13}$$

A trivial exception from this estimation would be an unusual static force field varying periodically with alternating signs versus distance. However, there is no indication to suppose/consider such strange fields at the moment.

These limitations are not directly applicable to the transient interaction model of Eq. 6 because in that case the energy density can be strongly inhomogeneous and the energy conservation law can be preserved even with a stronger transient interaction over a larger distance considering the slowness of this force.

### **2.3.** *Further extrapolations of the hypothesis*

When we talk about information, the question of coding is a natural one. If any one of the interactions outlined above exists one could seek for more interaction types between structures which can be transformed into each other by coding. Non-reversible coding would pose even more questions. Coding would have a role when there is a device between the two interacting structures and the interaction via the device would be modified which could be represented by coding operator $\bar{C}$ acting on the information. The correlation between the information patterns at the static interaction described in section 2.1 can symbolically be represented as:

$$I_{1,2} = \langle S_1 ; S_2 \rangle \tag{14}$$

where the angle bracket represents the correlation (mutual information, etc.) within the patterns $S_1$ and $S_2$. To generalize this notation, we can write for the "encoded" type of interaction that the correlation between the information patterns for this new type of interaction can symbolically represented as:

$$J_{1,2} = \left\langle S_1 ; \bar{C} S_2 \right\rangle \tag{15}$$

$$J_{1,2} = \left\langle \bar{C}_1 S_1 ; \bar{C}_2 S_2 \right\rangle \tag{16}$$

where the encoding operators $\bar{C}$, $\bar{C}_1$ and $\bar{C}_2$ represent/include the different kinds of situations, devices, related initial and boundary conditions, etc.

Going even further in this direction, proper $\bar{C}_t(t_1,t_2)$ encoding operators may interrelate structures in the past and future and another interaction represented by the correlation between present and future:

$$J_{t_1,t_2} = \left\langle \bar{C}(t_1,t_2) S_{t_1} ; S_{t_2} \right\rangle \quad (17)$$

where $S_{t_1}$ and $S_{t_2}$ are the structure and present time $t_2$ and its evolved version at time $t_2$ , respectively. Eq. 17 or similar forms may be the ways to new variation principles for describing the evolution of a system, for example, by assuming that the maxima or minima of $J_{t_1,t_2}$ select the possible paths of evolution of the system and/or the probability of a certain path is a monotonic function of $J_{t_1,t_2}$ . Obviously, we do not suppose any more that the interaction discussed here is based on direct energy exchange or that it directly involves an energy term.

Finally we would like to note that, even though these ideas may look very unusual or crazy, many known effects could already be envisioned as a realization of the encoded type of interaction, either energy-based or probability based. Without further details, some of the examples are: it seems to be possible to introduce simple coding operators for gravitation interaction, Coulomb interaction, and if we generalize toward continuum information measures, variation calculus, etc. If we consider quantum physical states/wavefuntions information patterns, the Pauli-principle of fermions and statistics of bosons may be interpreted as probability interactions with proper coding operators. The double slit experiment could be interpreted as evolution based on $\bar{C}_t(t_1,t_2)$ encoding operators. In this paper, by no means do we want to enter in a discussion about these matters. We only want to indicate that the seemingly strange hypotheses of Section 2 may be not so weird, after all.

## 3. Experiments

We used a HR-202i precision balance for the measurements that has a resolution of 10 microgram up to 40 grams of load. This is obviously an insufficient resolution for studying possible forces between two devices. Accuracies achieved at Eotvos-type or other pendulum experiments are enough to test gravitation interaction between small bodies and we are far from that sensitivity. Thus the only hope to see a static force effect was the possibility of a sharp "resonance"-like interaction between completely random large patterns that are identical.

The structural information patterns were zeroes, white noise with uniform distribution and Gaussian 1/f noise. The white noise data contain the most information, the 1/f noise contains less information due to logarithmically decaying correlations and zeroes contain zero information. The reason we tested also the 1/f noise is the fact that structural patterns of nature often contain 1/f noise and using the same kind of test source may "resonate" with natural patterns due to the length/time scale invariance of 1/f noise patterns. The 1/f noise records contained independent 256 kB long files with Gaussian 1/f noise.

We carried out experiments by using USB-2 flash drives (1÷16GB), DVD and CD disks to determine if such a force interaction is detectable at our resolution.

The temperature during the measurement was between 25 and 25.5 $^{o}$C (77÷78F) and the humidity air was in the range of 29 ÷ 46 %.

### 3.1. *Precautions and some of the possible artifacts*

The HR-202i precision scale showed a natural relaxation time of the order of a minute when a test object with mass of similar range was carefully placed on it.

This precision scale is very sensitive for stronger shocks and these typically cause the shifting of zero level. To reduce shocks, we used a handmade aluminum folia container to hold the flash drives and MP3 player and to place them on the scale. In this way we could avoid the zero shifting shocks. Sudden changes in the humidity would also cause zero shifts. However, calibration measurements indicated that after zero correction, the measured values were correct. We processed those measurement data where there was no significant change in the zero level and the humidity during the measurement.

Concerning other possible artifacts, the most important one is the observed negative weight transient due to the heating of the information storage element by the recording device. We compared this weight loss and found 100÷1000 times greater than the expected mass loss originating from the following effects:

a) Weight loss due to the Archimedes law and the increased volume of the thermally expanded device.

b) Weight loss due to the evaporation of condensed surface water. Assuming surfaces with known amount of condensed water, such as metals, silicon we arrived at the conclusion that such effect would also be insignificant.

Therefore, we found that there are two remaining possibilities to explain the negative weight transient in the classical way:

1. The assumption of hygroscopic material components that can absorb much more water. Such a phenomenon could especially be relevant for CDs and DVDs, which have large surface, and are known to be hygroscopic. The effect would be determined by the combination of stored water and air humidity and their relation to the sorption isotherm of the unknown hygroscopic components in the device. However, the variations of the observed times constants of negative weight relaxation seem to be inconsistent with a simple picture.

2. Circulation due to convection (convection-circulation) caused by the air heated by the

warm device and cooled by the walls of the confined space of the scale. Such a phenomenon could especially be relevant for Flash drives and MP3 players, which have small surface, and elevated temperature. The resulting flow causes a lifting Bernoulli force. This effect certainly contributes to the observed transient effects, though it implies the same relaxation time constant for all cases. See the picture in Section 4.

### **3.2.** *Results*

The summary of the results is as follows.

i) Due to the non-reproducible changes of static weight of these memories after changing the information in them, we have not been able to clarify the existence of a reproducible static weight change that could be interpreted as a force interaction between the memory and its environment, even though the observed large variations of static weight (sometimes up to the milligram range at weather change) could possibly be the manifestation of the new type of interaction with changing environment.

ii) Interaction between the memories with the same information was not detected at the given weight resolution. In the experiments one flash drive filled up with white noise was placed on the scale and its weight was determined. Then another flash drive with the same information content was placed over it at a distance of 1 cm. Possible changes in the weight of the first flash drive were then monitored. We typically observed a small weight transient (in the order of 0.03) mg, which gradually relaxed toward zero within a time period of a minute, and we interpreted this effect as an artifact caused by the thermal disturbance of the scale's internal atmosphere when we opened its door and put the other flash drive there. The experiment with the largest information content was done with the combination of a 2GB and a 16GB flash drive filled by 7.5 times (15GB) of the same information as in the 2GB drive. Using the negative result and our resolution in the last term of Eq. 5 we obtain

$$10^{-7} N > \frac{\eta P}{(10^{-2} m)^2} 7.5 * (8 * 2 * 10^9 bit)^2 \quad . \tag{18}$$

Thus our result for the $\eta$ constant and the pattern-specific $P$ parameter is as follows:

$$\eta P < 5.21 * 10^{-33} \frac{N}{bit^2 m^2} \tag{19}$$

iii) The hypothesis of slow transient interaction (Eq. 6) is more in line with the observations provided the observed effects are not artifacts. Originally we expected only a static interaction but due to the spurious experimental results we expanded the picture with the transient effect idea. Therefore, if further studies with different scales and better control of experimental conditions provide a classical explanation for the transient weight effects, the transient picture should be removed from the theory and the related parts of the present paper should be disregarded. However, if the information-dependence of the transient force is confirmed, then it may mean that writing or deleting the information in any of these devices causes peculiar negative weight transients, up to mass fraction of about $10^{-5}$, which is followed by various types of relaxation processes. These relaxations have significantly different dynamics compared to transients observed during cooling after stationary external heating. Interestingly, a USB-1 MP3 player has also suffered comparable mass loss during playing music, even though its information content was not changed, and its power dissipation and warming-up was much less than that of the USB-2 flash drives. As we mentioned above, classical interpretation of the negative weight transients could be absorbed water in hygroscopic components or lifting Bernoulli force due to convection of warm air, however comparison of relaxation time constants at the various measurements does not yet support an obvious explanation in this fashion. However note: no comparison with passive thermal relaxation has been carried out at MP3 players yet.

Now, let us see the specific results. In Figure 1, the negative weight transient and static weight loss of a CD-R disk (Verbatim) after burning 610 MB white noise on it. Note: CDs are known to be somewhat hygroscopic.

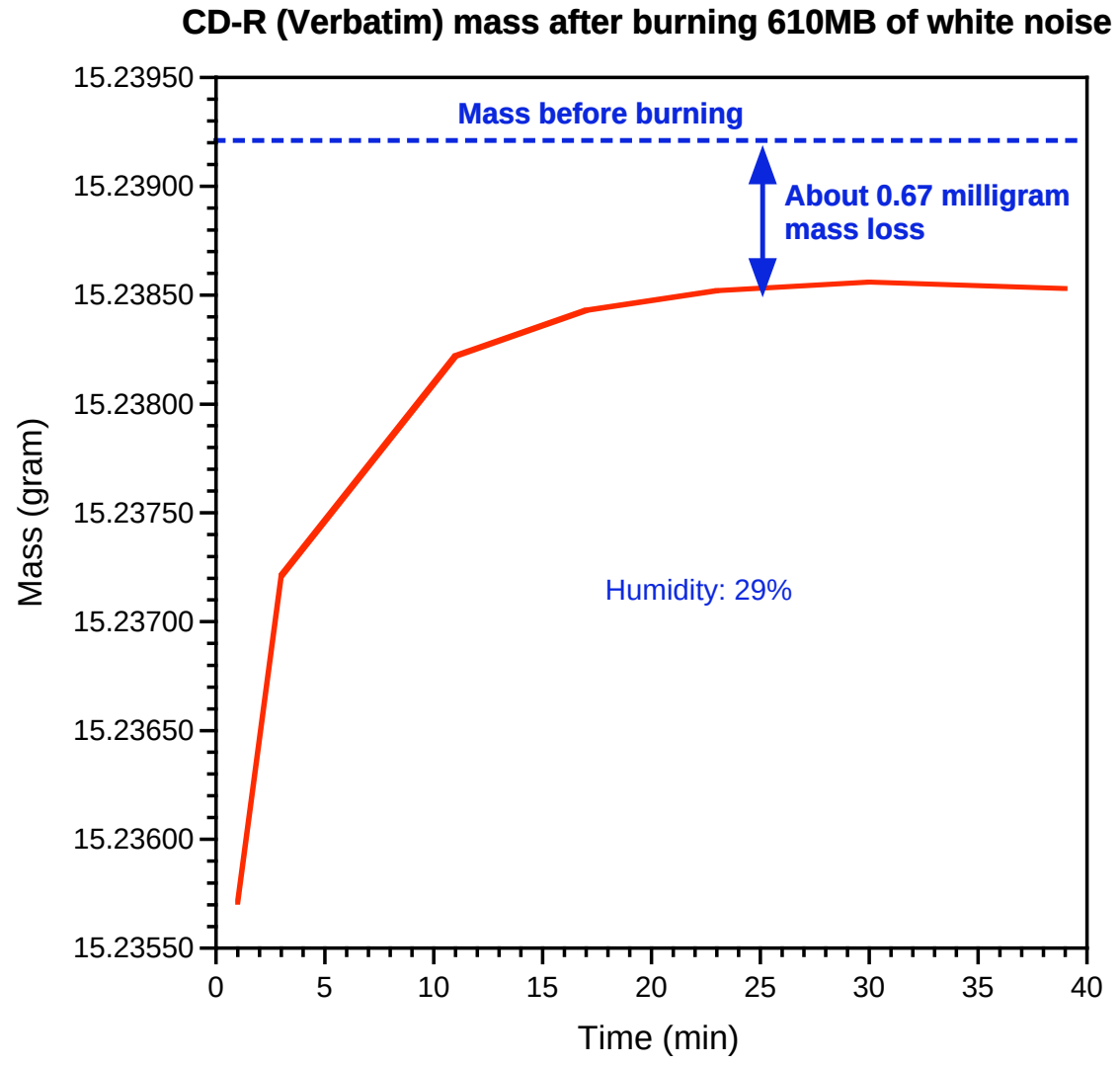

**Figure 1.** Negative weight transient and static weight loss of a CD-R disk (Verbatim) after burning 610 MB white noise on it. Note: CDs are known to be somewhat hygroscopic.

In Figure 2, the semi logarithmic plot of the absolute value of the negative mass transient in Figure 1 is shown. A straight line represents an exponential decay. Solid line: measurements; dashed line: exponential fit. The relaxation time constant is about 5.5 minute and that is much longer than the estimated thermal relaxation time (0.5÷1 minute).

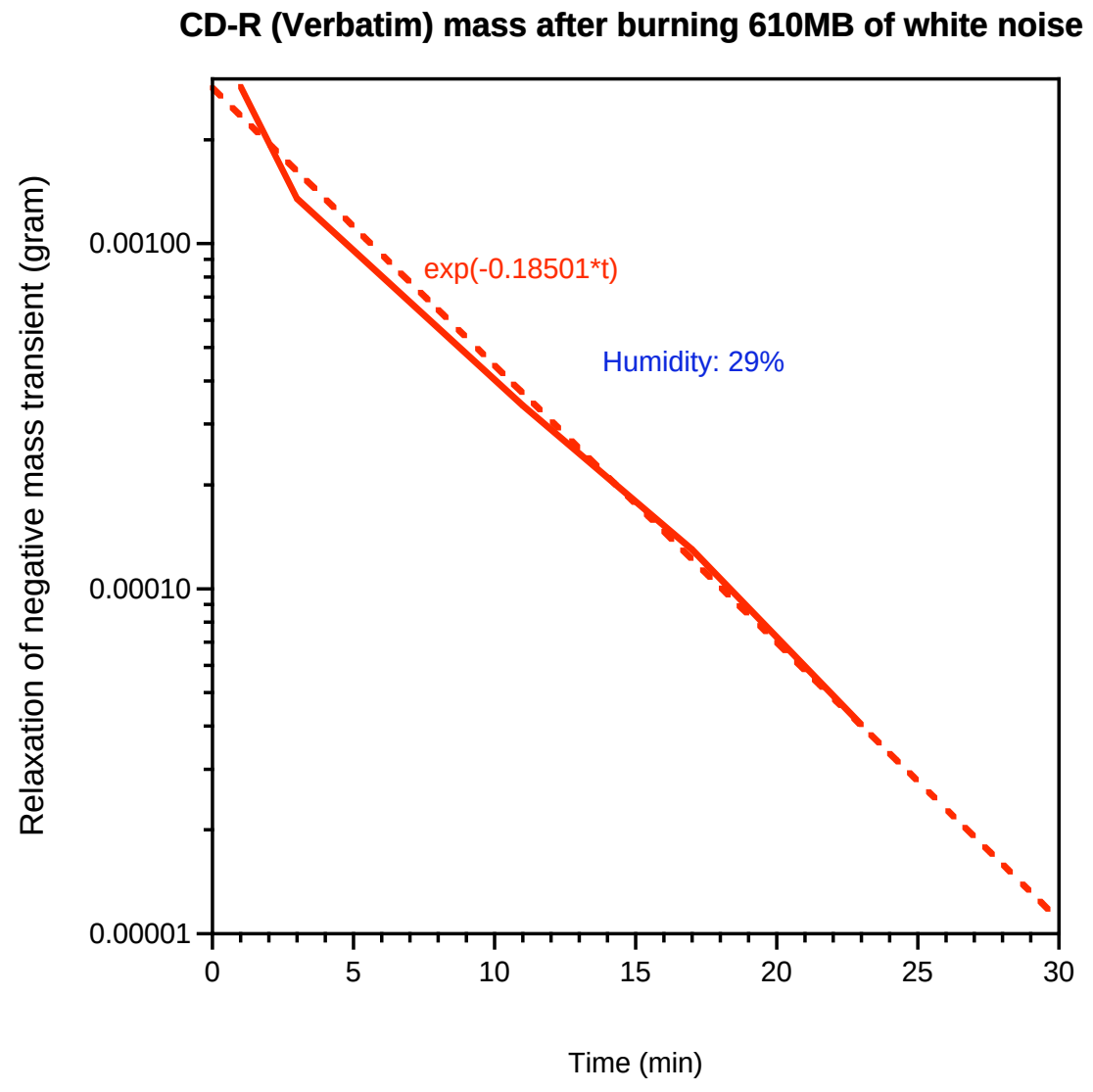

**Figure 2.** Semi logarithmic plot of the absolute value of the negative mass transient in Figure 1 is shown. A straight line represents an exponential decay. Solid line: measurements; dashed line: exponential fit. The relaxation time constant is about 5.5 minute and that is much longer than the estimated thermal relaxation time (0.5-1 minute).

In Figure 3, the negative weight transient and static weight loss of a DVD-R disk (Maxell) after burning 4.26 GB white noise on it. Note: DVDs are known to be about 10 times less hygroscopic than CDs.

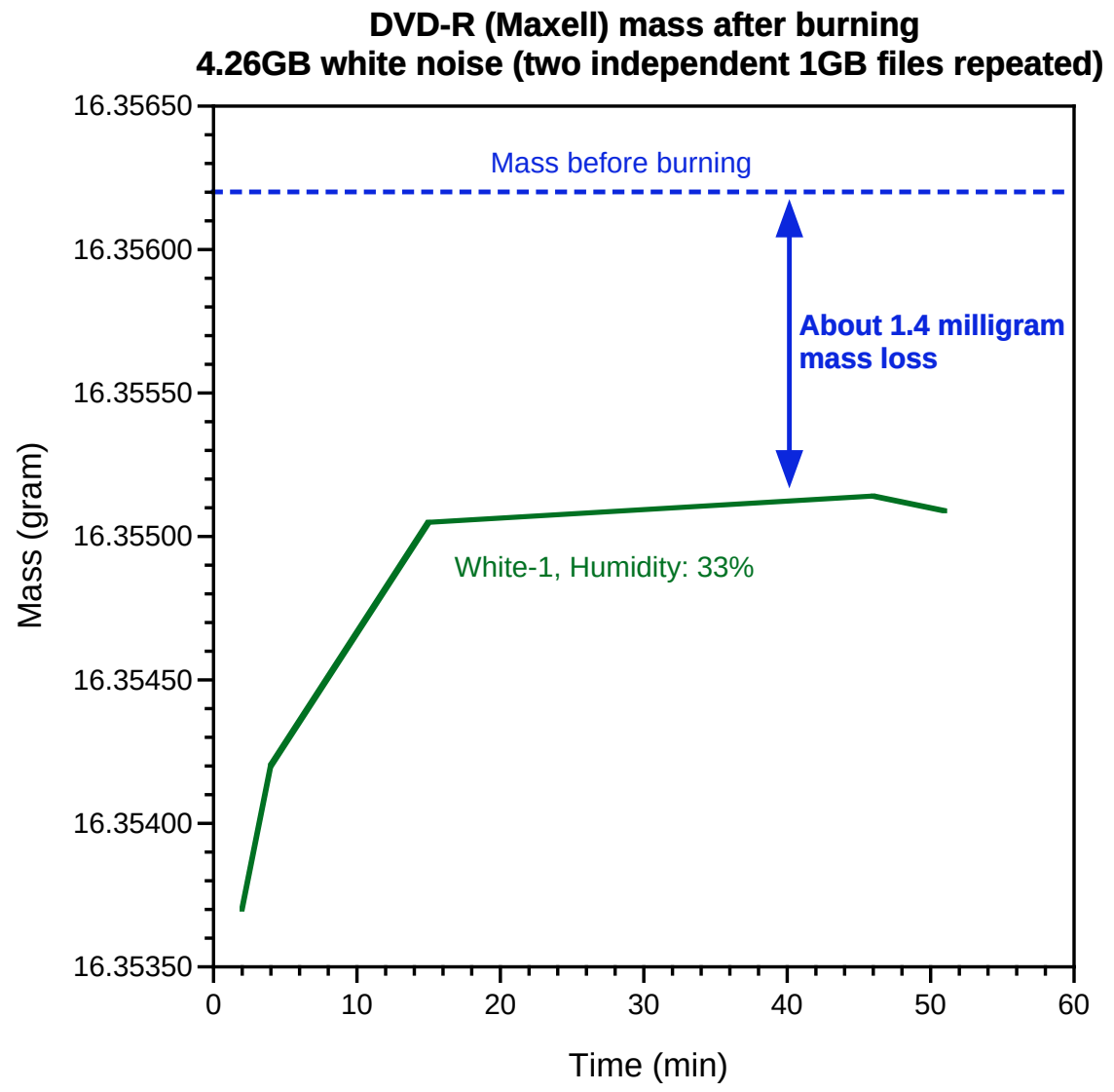

**Figure 3.** Negative weight transient and static weight loss of a DVD-R disk (Maxell) after burning 4.26 GB white noise on it. Note: DVDs are known to be about 10 times less hygroscopic than CDs.

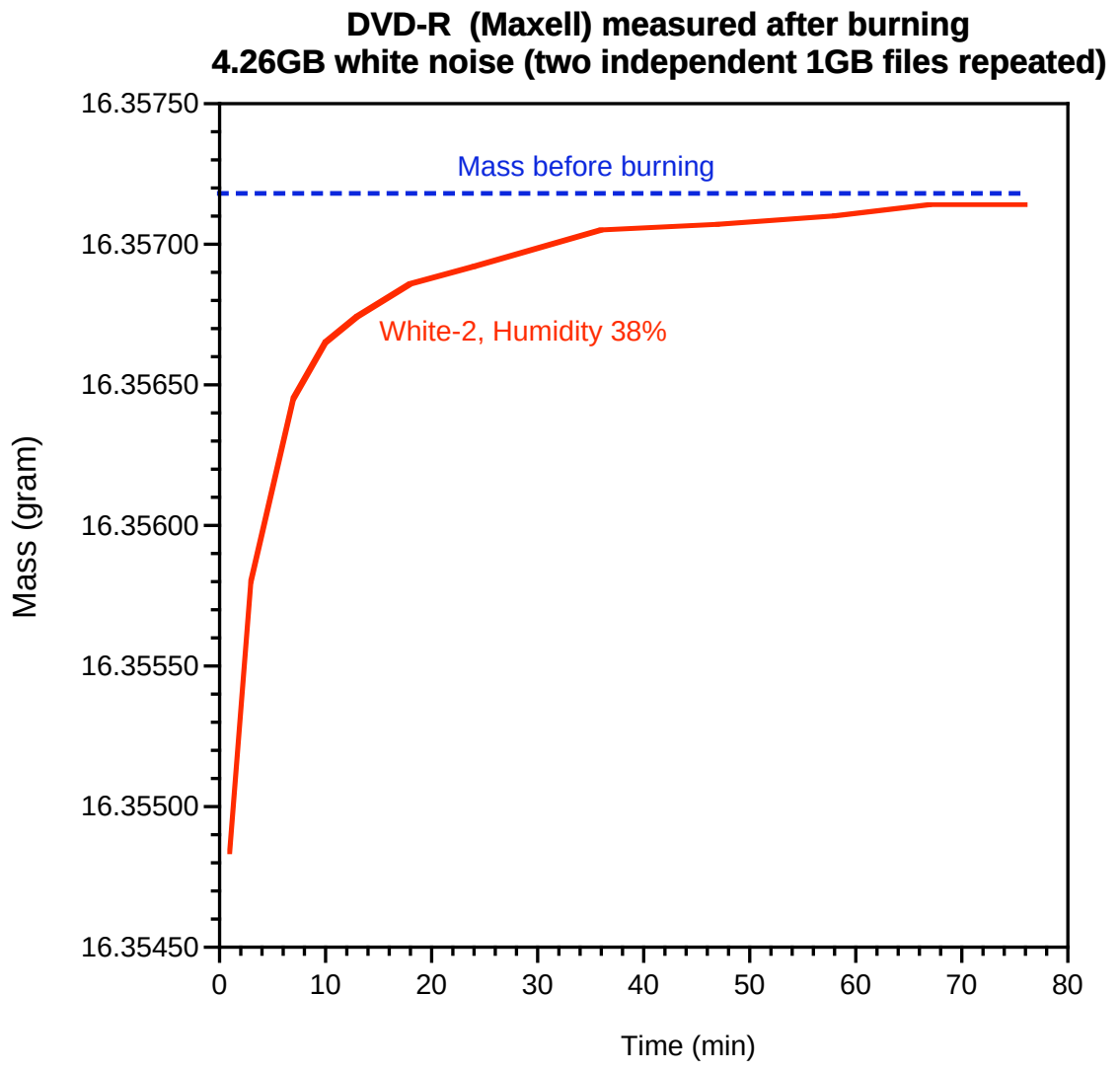

**Figure 4.** Negative weight transient and static weight loss of another DVD-R disk (Maxell), at higher humidity, after burning 4.26 GB white noise on it. The static weight change is significantly less.

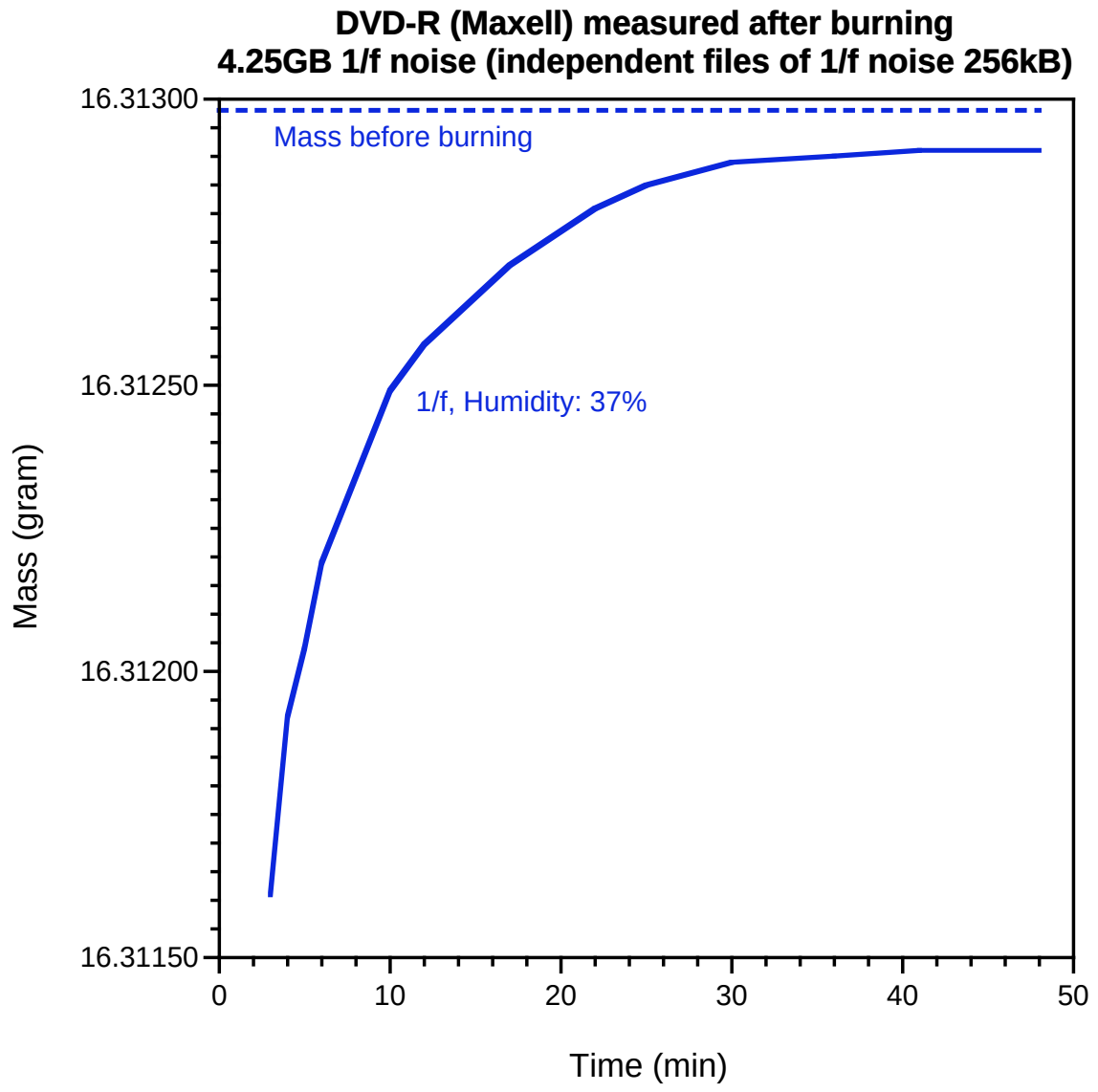

**Figure 5.** Negative weight transient and static weight loss of another DVD-R disk (Maxell), also at higher humidity, after burning 4.25 GB 1/f noise on it. The static weight change is also significantly less than in the case of the first DVD.

In Figure 6, the semi logarithmic plot of the absolute value of the negative mass transients in Figures 3-5 are shown for comparison. A straight line represents an exponential decay. Solid line: measurements; dashed line: exponential fit. The shortest relaxation time constant is about 5 minute and even that is much longer than the estimated thermal relaxation time (0.5-1 minute). The longest relaxation time constant is about 15 minutes. These data contradict to a simple interpretation based on hygroscopic materials and air humidity. The expectations based on this simple picture are the opposite to what we see: the higher the humidity of air the faster the re-absorption of lost water and the shorter the mass relaxation time should be.

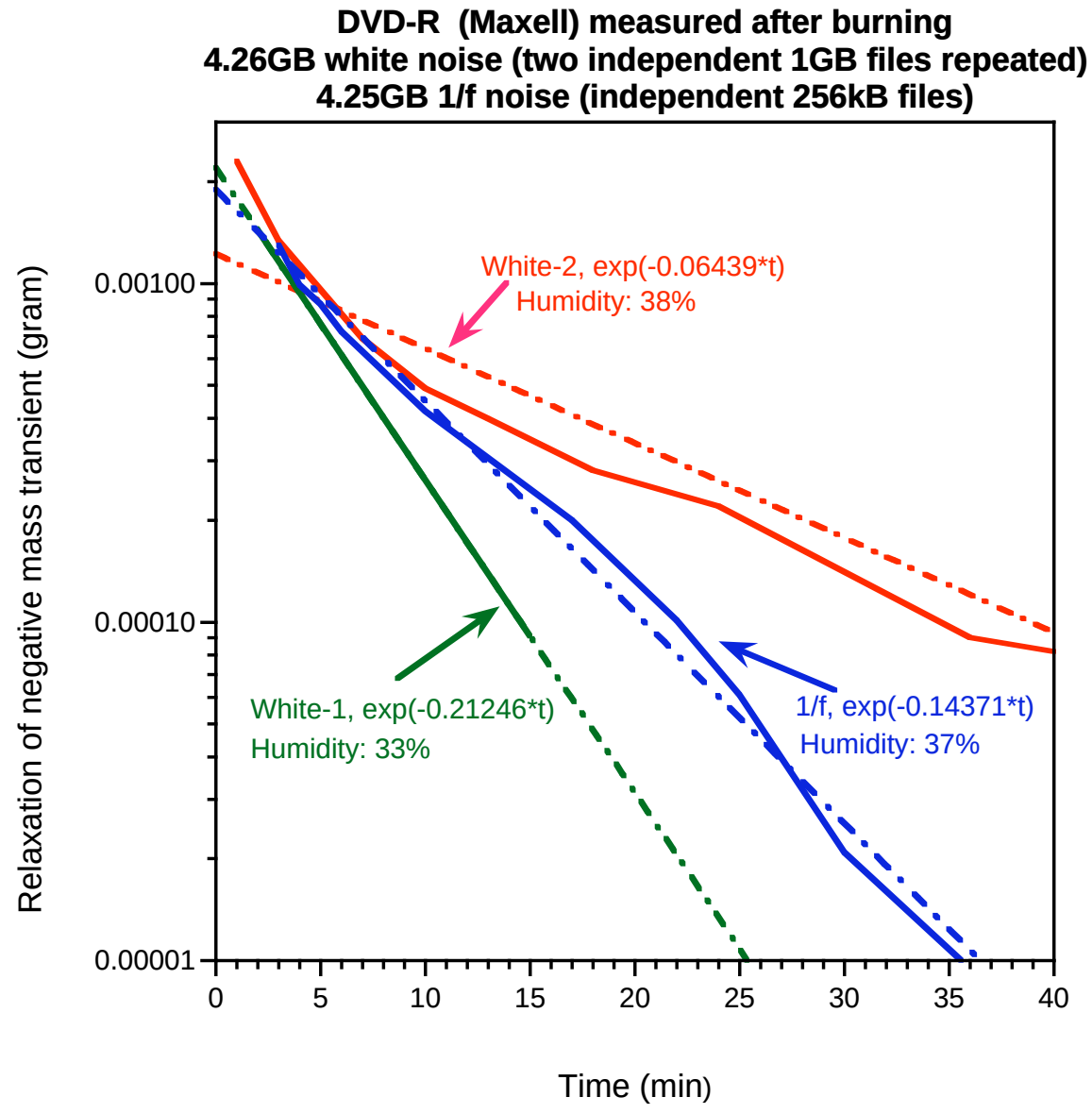

**Figure 6.** Semi logarithmic plot of the absolute value of the negative mass transients in Figures 3-5 are shown for comparison. A straight line represents an exponential decay. Solid line: measurements; dashed line: exponential fit. The shortest relaxation time constant is about 5 minute and even that is much longer than the estimated thermal relaxation time (0.5-1 minute). The longest relaxation time constant is about 15 minutes.

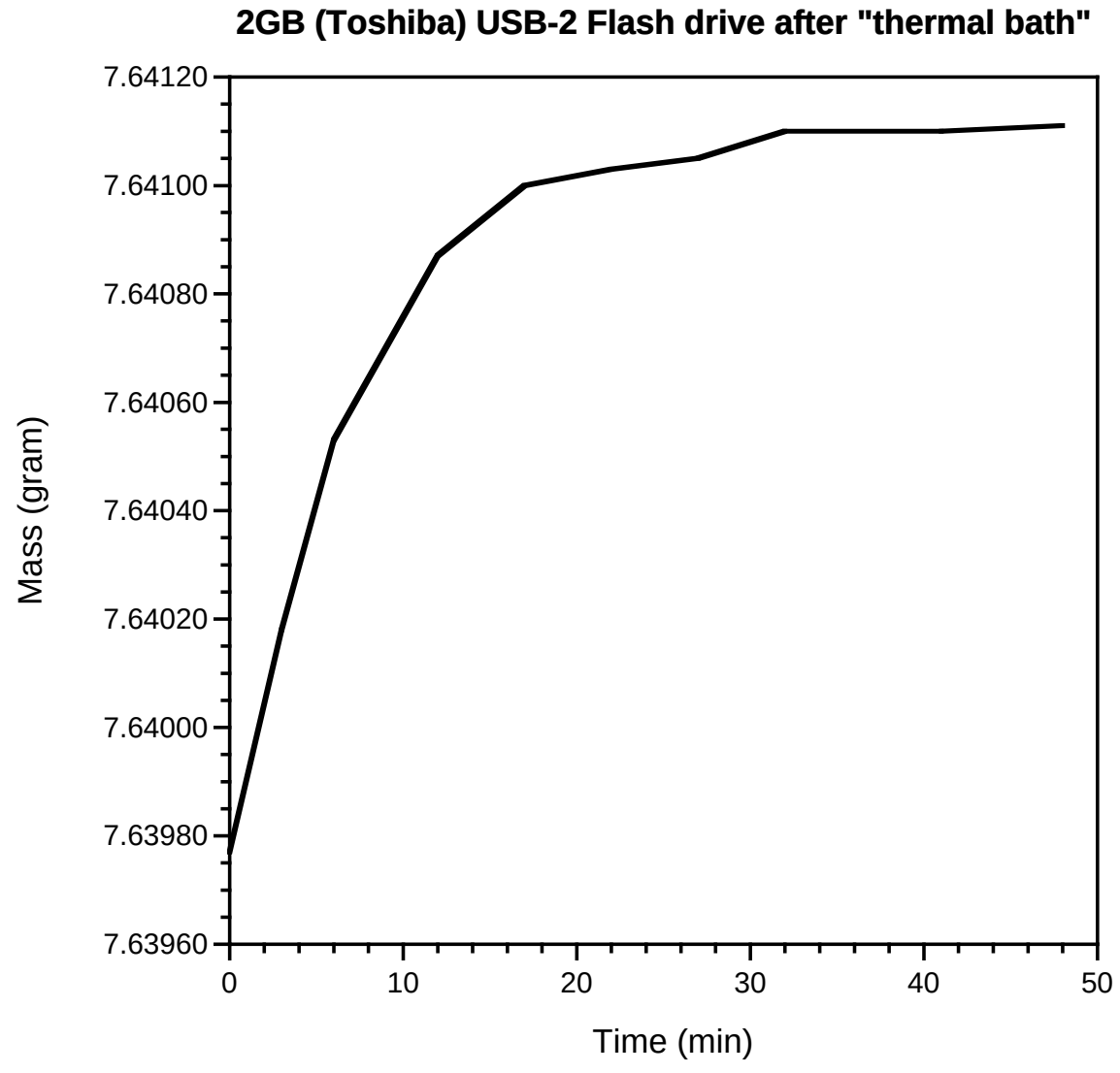

**Figure 7.** Negative weight transient of a 2GB USB-2 flash drive (Toshiba) after a thermal bath.

In Figure 7, the negative weight transient of a 2GB USB-2 flash drive (Toshiba) after a thermal bath is shown. The thermal bath was a pint of hot water in which the flash drive was immersed after including in a triple layer of plastic bags. A thermometer was placed at the flash drive. The initial temperature was 73 °C; ten minutes later 64 °C; and 30 minutes later 43 °C; when the flash drive was removed from the thermal bath and was put on the scale to monitor its weight evolution. The evolution of the mass indicates that during the information exchange, related heating effects are certainly a factor, either due to the convection-circulation; the water-absorption; or other mechanism. However, we will see that the actual measurements with changing information in the device indicate a different behavior with longer relaxation times.

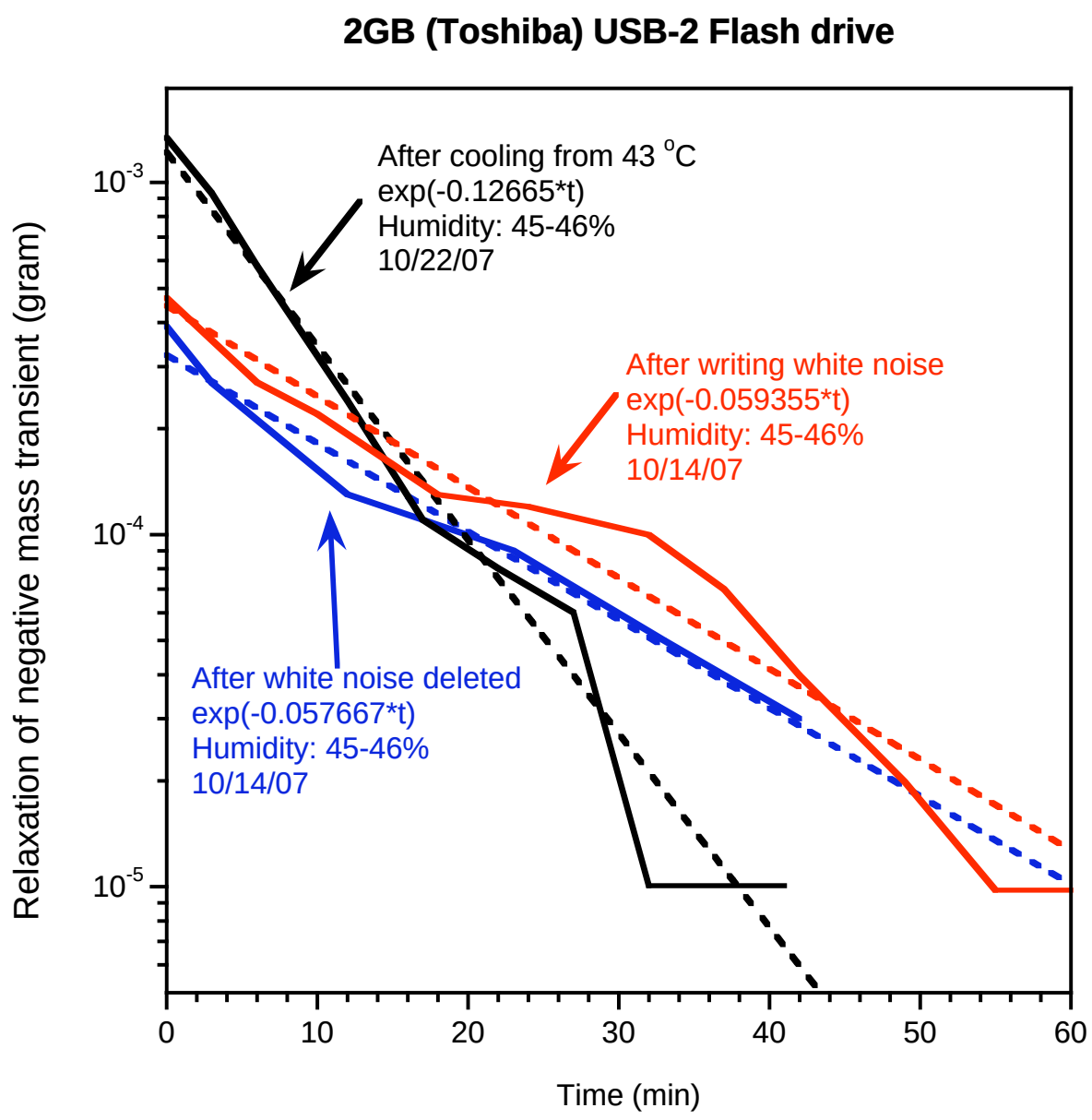

**Figure 8.** Semi logarithmic plot of the absolute value of the negative mass transients measured after the thermal bath (see Fig. 7) and after information change at different conditions are shown for comparison. A straight line represents an exponential decay. Solid line: measurements; dashed line: exponential fit. The relaxation time constant after the thermal bath is about half of the relaxation time observed after recording or deleting white noise from the device.

Figure 8 shows a semi-logarithmic plot of the absolute value of the negative mass transients measured after the thermal bath (see Fig. 7), and that of after information change at different conditions. A straight line represents an exponential decay. Solid line: measurements; dashed line: exponential fit. The relaxation time constant after the thermal bath is about half of the relaxation time observed after recording or deleting white noise from the device. The interpretation by the convection-circulation would predict the same time constant for each case. The interpretation based on hygroscopic materials could work if we suppose that different hygroscopic parts are dominant after thermal bath than after information change because of local hot spots in the device during information recording. However the small surface and the large weight change makes also this explanation questionable.

In Figure 9, a semi logarithmic plot of the absolute value of the negative mass transients measured in a 16 GB USB-2 Flash drive (Axiom) after information change with different conditions are shown for comparison. A straight line represents an exponential decay. Solid line: measurements; dashed line: exponential fit. The 1/f noise experiments are not conclusive because it takes 7 hours to fill the drive with 1/f noise and the filling was done during night and relaxation experiments started only hours later. Even the other experiments have speed problems because to delete the information or to fill up the drive with white noise takes more than 1 hour. In any case, the relaxation times are in a qualitative agreement with the air humidity even though it is not clear if such small variations of the humidity could influence the mass relaxation time constant so significantly and if the small surface and large weight change could be justified with the materials used in the device.

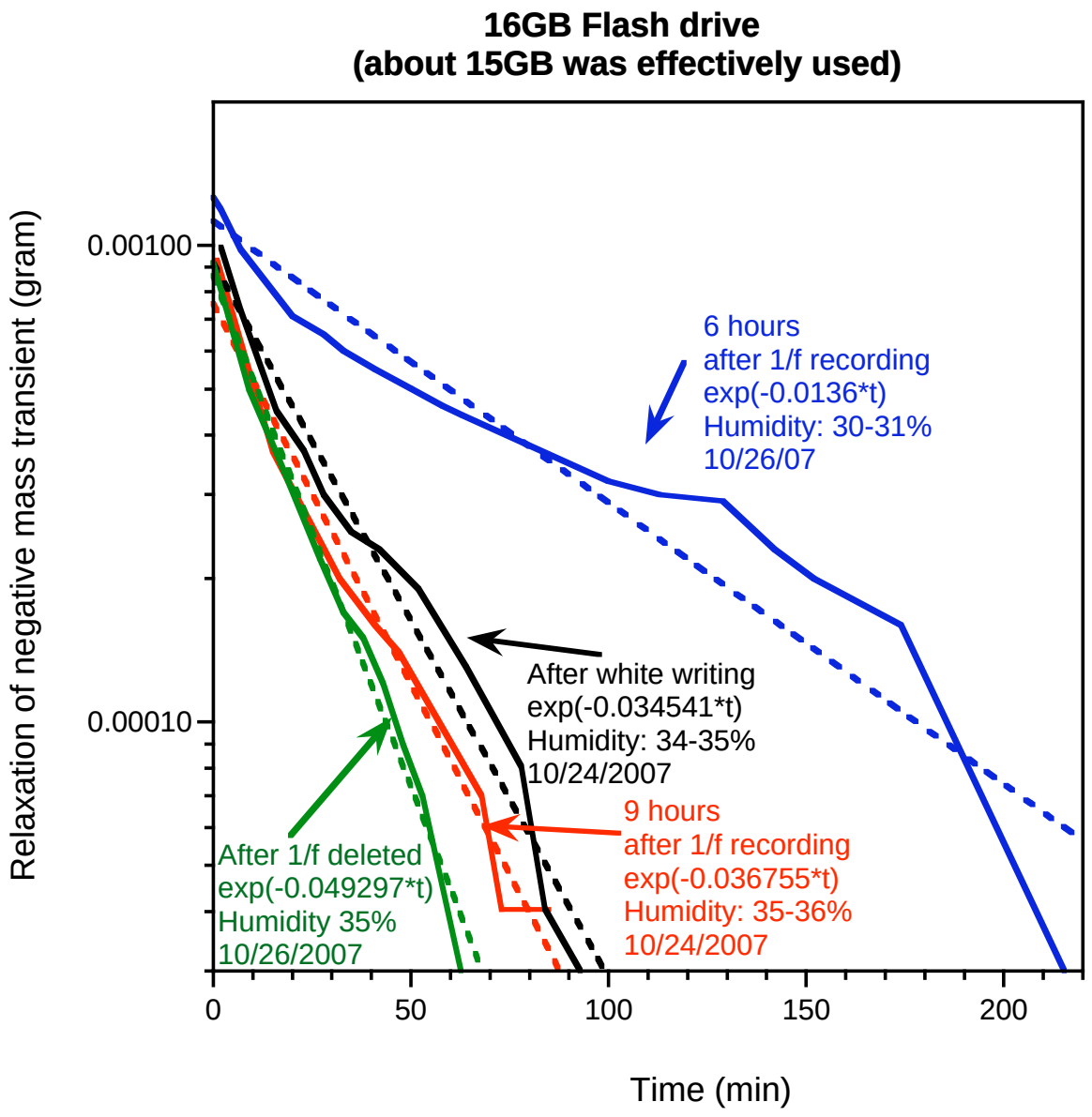

**Figure 9.** Semi-logarithmic plot of the absolute value of the negative mass transients measured in a 16 GB USB-2 Flash drive (Axiom) after information change with different conditions are shown for comparison. A straight line represents an exponential decay. Solid line: measurements; dashed line: exponential fit. The 1/f noise experiments are not conclusive because it takes 7 hours to fill the drive with 1/f noise. The dates exclude the possibility of different relaxation time constant due to an aging effect.

Finally we considered the question if storing a new information is needed or rather the generation or measurement of that information, that is an information flow, is enough to cause weight reduction in such devices. Because of their 10 times smaller power dissipation than that of USB-2 flash drives, we selected a USB-1 MP3 player to generate information. In Figure 10 a semi-logarithmic plot of the absolute value of the negative mass transients measured in a uTronix OT403 512MB MP3 Player (and USB Flash Drive) after the playback of music (see the Appendix) for 40 minutes. The player gradually lost 1.14 milligram of weight during the playback and this figure shows the relaxation backward up to the saturation point where the weight grew less than our resolution (10 microgram) over 10 minutes. We show the first result here and note that passive thermal relaxation experiments would confirm if there is any difference as it is with Flash drives.

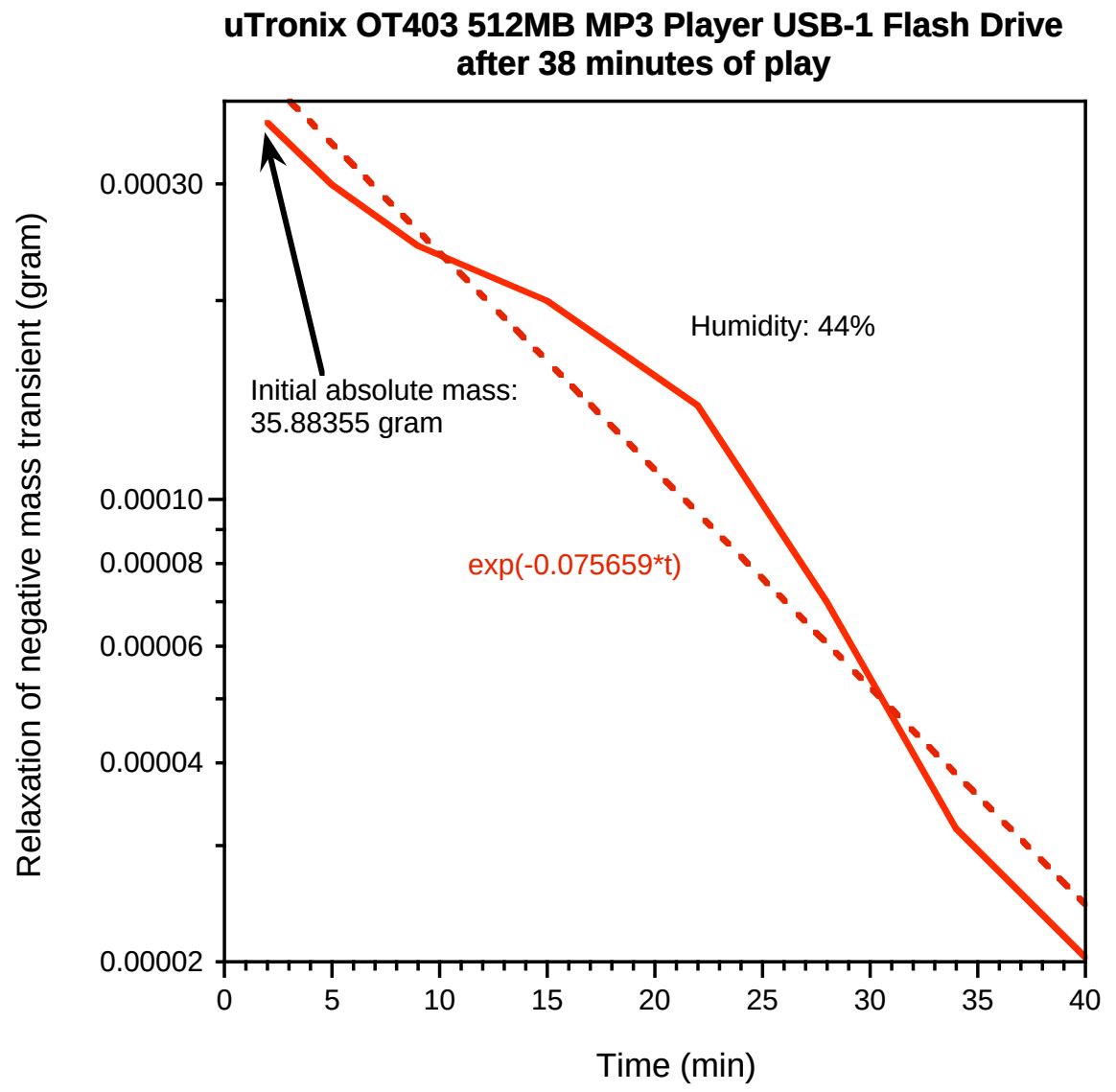

**Figure 10.** Semi-logarithmic plot of the absolute value of the negative mass transients measured in a uTronix OT403 512MB MP3 Player (and USB Flash Drive) after the playback of music (see the Appendix) for 40 minutes. The player gradually lost 1.14 milligram of weight during the playback and this figure shows the relaxation backward up to the saturation point where the weight grew less than our resolution (10 microgram) over 10 minutes.

## 4. Summary

We hypothesized possible new types of forces that would be the result of new types of interactions, static and a slow transient, between objects with related information contents (pattern). We first supposed a static force however, due to the unexplained experimental effects of slow transients, we later expanded the hypothesis with a possibly additive transient force that shows a slow relaxation. Any of these mechanisms could make material composition dependence claimed by Fishbach, *et al.* in Eotvos type experiments plausible. We carried out experiments by using a high-resolution scale with the following memories: USB-2 flash drives (1-16GB), DVD and CD disks to determine if such an interaction exist/detectable with a scale resolution of 10 microgram with these test objects. We applied zero information, white noise and 1/f noise type data. Due to the non-reproducible changes of static weight of these memories after changing the information in them, we have not been able to clarify the existence of a reproducible static force between the memory and its environment even though the variations of static weight could possibly be the manifestation of the new type of interaction with changing environment. Interaction between two memories containing the same information was not detected at the given weight resolution. The hypothesis of slow transient interaction is more in line with the observations provided the observed effects are not artifacts. Writing or deleting the information in any of these devices causes peculiar negative weight transients, up to milligrams (mass fraction around $10^{-5}$), which is followed by various types of long-time relaxation processes. These relaxations have significantly different dynamics compared to transients observed during cooling after stationary external heating. Interestingly, a USB-1 MP3 player has also developed comparable transient mass loss during playing music, even though its information content was not changed, and its power dissipation and warming-up was much less than that of the USB-2 flash drives. A classical interpretation of the negative weight transients could be absorbed water in hygroscopic components, though comparison of relaxation time constants with air humidity data does not support an obvious explanation. Another classical interpretation with certain contribution is the lifting Bernoulli force caused by the circulation due to convection of the warm air, see Figure 11.

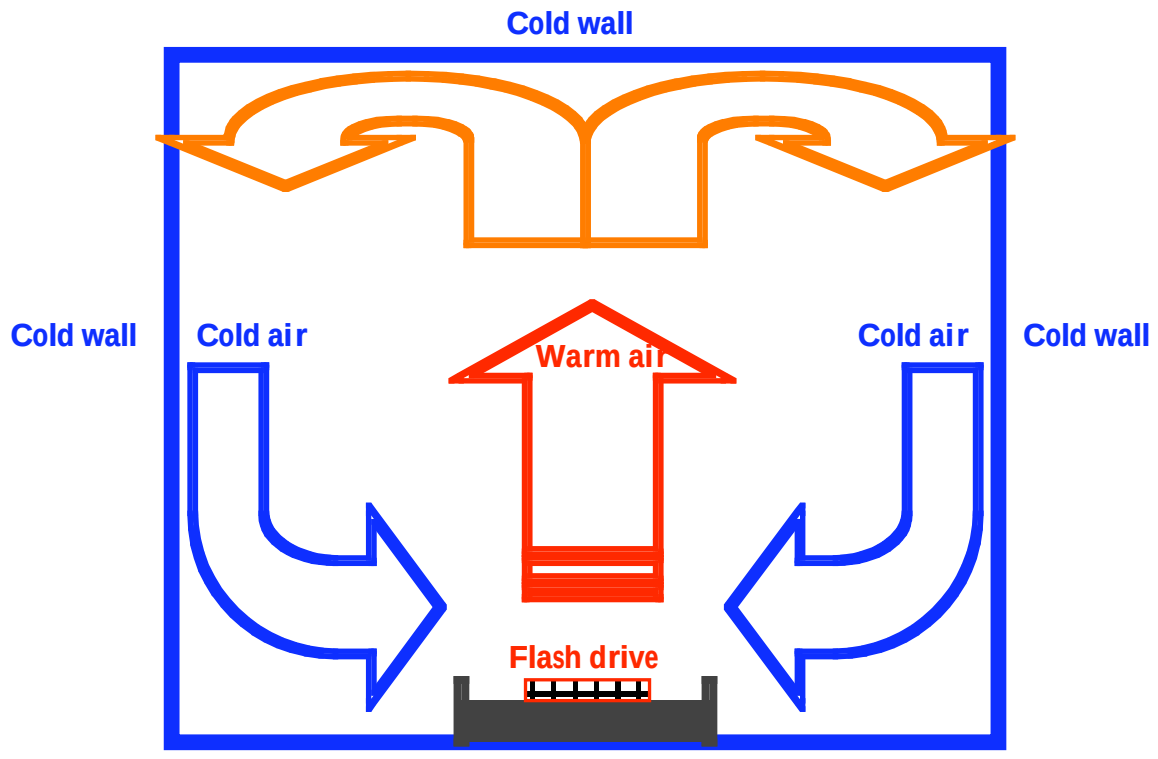

**Figure 11.** Circulation due to convection caused by the air warmed by the Flash drive and cooling by the walls of the confined space of the scale. The resulting flow causes a lifting Bernoulli force. This effect certainly contributes to the observed transient effects, though it implies the same relaxation time constant for all cases.

However, in this case all observed time constant should have been the same unless some hidden parameter causes the observed variations. Thus further studies are warranted to clarify if there is indeed a new force, which is showing up as negative mass at weight measurement when high-density structural information is changed or read out (measured).

Just for curiosity, we estimate (see in the Appendix) that, if the information-based interaction is real and scales with information handling performance, the weight of bodies might be reduced and floating engines might be achieved with several orders of magnitude greater data handling capacities than today's values at the same mass. However, this assumption is yet a science fiction and it may stay like that.

Finally, we would like to emphasize that even if the transient weight behavior will ultimately turn out to be an artifact, the question of the existence of interaction between information patterns will remain open in both the direct form and the coded form, respectively, see the considerations about the static interaction in Sections 2.1 and 2.3.

## 5. Acknowledgements

A discussion with Krishna Krishna Narayanan and Arun Srinivasa (about the nature of information 2007); Kazuaki Kuroda (about gravitation measurements 1991); and Sergey Bezrukov (potential problems with the interpretation of weight transients for a biological system 2007) are appreciated.

## Appendix A. If the unexpected got confirmed: floating devices and vehicles

Until origin of the observed negative mass transients is clarified, the following considerations should be considered as something that will stay as a science fiction with a high probability. If the unexpected gets confirmed and information change/flow indeed causes a repulsive interaction, which can be used to decrease gravitation effects, one can envision devices and vehicles that are able to overcome gravitation force. The key parts of such equipment are shown in Fig. 12. The acceleration sensor produces an error signal, which is used by the control loop to achieve the required acceleration. The noise pattern and information flow speed is controlled to match local conditions of environmental information patterns and actual requirements of vertical acceleration/deceleration.

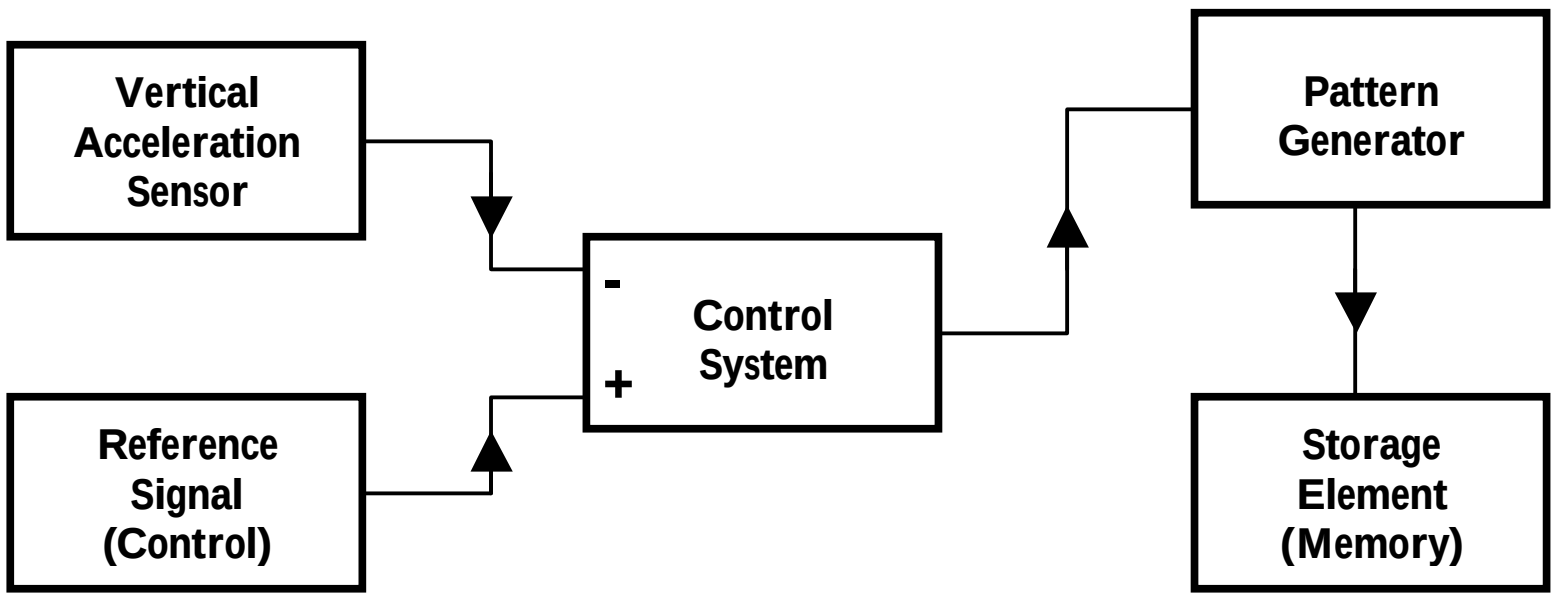

**Figure 12.** Key elements of a floating device if the repulsive force effect is real. This is yet a science fiction and it may stay like that.

## Appendix B. List of music during the buildup of the mass transient in the USB-1 MP3 player

The list of music played in the MP3 player during the mass reduction was as follows; Crosby and Nash: Immigration man; Metallica: Nothing else matters; Sher: Gypsies, Trams, Thieves; Creedence Clearwater Revival: Who'll stop the rain; Wishbone Ash: Errors of my way; Brian Eno: Final sunset; Roger Daltrey: Giving it all away; King Crimson: The talking drum, Lark's tongues in aspic 1 (fraction).